\documentclass[10pt,conference,letterpaper]{IEEEtran}

\usepackage{xcolor}
\usepackage{acronym}
\usepackage{caption}
\usepackage{subfig}
\usepackage{numdef}
\usepackage{numprint}
\usepackage{longtable}
\usepackage{multirow}
\usepackage{booktabs}

\usepackage{float}

\newcommand{\specialcell}[2][c]{%
  \begin{tabular}[#1]{@{}c@{}}#2\end{tabular}}

\newcommand{\ie}{i.e.,}
\newcommand{\eg}{e.g.,}

\usepackage{flafter}

\usepackage[export]{adjustbox}

\usepackage{array}
\newcolumntype{L}[1]{>{\raggedright\let\newline\\\arraybackslash\hspace{0pt}}m{#1}}
\newcolumntype{C}[1]{>{\centering\let\newline\\\arraybackslash\hspace{0pt}}m{#1}}
\newcolumntype{R}[1]{>{\raggedleft\let\newline\\\arraybackslash\hspace{0pt}}m{#1}}

\usepackage{enumitem}
\newlist{MyIndentedList}{itemize}{4}
\setlist[MyIndentedList,1]{%
    label={},
    noitemsep,
    leftmargin=0pt,
    }
\setlist[MyIndentedList]{%
    label={},
    noitemsep,
    }

\begin{document}

\title{
Adapting Software Architectures to \\ Machine Learning Challenges
{
}}

\author{\IEEEauthorblockN{Alex Serban}
\IEEEauthorblockA{
\textit{iCIS, Radboud University}\\
The Netherlands \\
a.serban@cs.ru.nl}
\and
\IEEEauthorblockN{Joost Visser}
\IEEEauthorblockA{
\textit{LIACS, Leiden University}\\
The Netherlands \\
j.m.w.visser@liacs.leidenuniv.nl
}
}

\maketitle

\begin{abstract}
Unique developmental and operational characteristics of \ac{ML} components as well as their inherent uncertainty
demand robust engineering principles are used to ensure their quality.
We aim to determine how software systems can be (re-) architected to enable robust integration of \ac{ML} components.
Towards this goal, we conducted a mixed-methods empirical study consisting of (i) a systematic literature review to identify the challenges and their solutions in software architecture for \ac{ML}, (ii) semi-structured interviews with practitioners to qualitatively complement the initial findings and (iii) a survey to quantitatively validate the challenges and their solutions.
We compiled and validated twenty challenges and solutions for (re-) architecting systems with \ac{ML} components.
Our results indicate, for example, that traditional software architecture challenges  (\eg~component coupling) also play an important role when using \ac{ML} components; along with new \ac{ML} specific challenges (\eg~the need for continuous retraining).
Moreover, the results indicate that \ac{ML} heightened decision drivers, such as privacy, play a marginal role compared to traditional decision drivers, such as scalability.
Using the survey we were able to establish a link between architectural solutions and software quality attributes, which enabled us to provide twenty architectural tactics used to satisfy individual 
quality requirements of systems with \ac{ML} components.
Altogether, the results of the study can be interpreted as an empirical framework that supports 
the process of (re-) architecting  software systems with \ac{ML} components.

\end{abstract}


\acrodef{ML}[ML]{machine learning}
\acrodef{AI}[AI]{artificial intelligence}
\acrodef{RF}[RF]{random forest}
\acrodef{DL}[DL]{deep learning}
\acrodef{NN}[NN]{neural networks}
\acrodef{DNN}[DNN]{deep neural network}
\acrodef{SE}[SE]{Software engineering}
\acrodef{SA}[SA]{Software architecture}

\acrodef{SLR}[SLR]{systematic literature review}
\newcommand{\allDocs}{\numprint{4200}}
\newcommand{\filteredDocs}{\numprint{2613}}
\newcommand{\manualInspection}{\numprint{82}}
\newcommand{\manualWithoutSnowbal}{\numprint{66}}
\newcommand{\snowballAll}{16}
\newcommand{\snowball}{13}

\newcommand{\intBeforeSnowball}{\numprint{29}}
\newcommand{\intRelevantArticles}{\numprint{42}}

\newcommand{\intInitialChallenges}{\numprint{18}}
\newcommand{\intInterviewChallenges}{\numprint{2}}
\newcommand{\intInterviewSolsolutions}{\numprint{0}}
\newcommand{\intAllChalenges}{\numprint{20}}

\newcommand{\intPercSLR}{\numprint{52}}
\newcommand{\intInterviewsPerc}{\numprint{46}}
\newcommand{\intInterviews}{\numprint{28}}

\newcommand{\intParticipants}{\numprint{10}}
\newcommand{\intCompanies}{\numprint{10}}
\newcommand{\intQuestions}{\numprint{31}}

\newcommand{\intVideo}{\numprint{8}}
\newcommand{\intMail}{\numprint{2}}
\newcommand{\finalQuestions}{\numprint{3}}
\newcommand{\removedQuestions}{\numprint{2}}
\newcommand{\intSections}{\numprint{5}}

\newcommand{\intQualityAtributes}{\numprint{14}}

\newcommand{\intMails}{\numprint{286}}

\newcommand{\intSurveyAll}{\numprint{53}}
\newcommand{\intSurveyParticipants}{\numprint{48}}

\newcommand{\intTechOrgPercentage}{\numprint{57}}
\newcommand{\intNonTechOrgPercentage}{\numprint{28}}
\newcommand{\intGovOrgPercentage}{\numprint{9}}
\newcommand{\intResearchOrgPercentage}{\numprint{6}}

\newcommand{\intExpbeg}{\numprint{13}}
\newcommand{\intExptwoyears}{\numprint{28}}
\newcommand{\intExpfiveyears}{\numprint{40}}
\newcommand{\intExpnine}{\numprint{19}}

\newcommand{\intEurope}{\numprint{53}}
\newcommand{\intUSA}{\numprint{34}}
\newcommand{\intAsia}{\numprint{13}}

\newcommand{\intTeamNine}{\numprint{43}}
\newcommand{\intTeamFifteen}{\numprint{34}}

\newcommand{\intCorr}{\numprint{210}}




\section{Introduction}
\label{sec:intro}

\ac{SA} plays an important role in data intensive systems, such as big data and analytics platforms.
However, until recently, the focus has been on the architectural decisions related to handling and storing large amounts of data and  on decisions that mitigate performance demands of analytics platforms~\cite{begoli2012design,sena2017characterizing}.

The interest to develop software with \acf{ML} components shifts the focus to decisions regarding the operational requirements of serving, monitoring, retraining and redeploying models~\cite{sculley2015hidden}. 
These decisions align with proposals to emphasise the operational aspect of \ac{SA}~\cite{woods2016operational}.
Moreover, the inherent uncertainty of \ac{ML} components demands a stronger emphasis on the uncertainty aspect of \ac{SA}; where the focus is on assessing the impact of uncertainty and  on the decisions made for its mitigation~\cite{serban2020towards,esfahani2013uncertainty}.

Although a significant body of literature studied the relevance of \ac{SA} for big data and analytics platforms~\cite{sena2017characterizing,avci2020software}, there is little empirical research on the role of \ac{SA} in systems with \ac{ML} components~\cite{lwakatare2019taxonomy,wan2019does}.
Our aim is to determine how software systems can be (re-) architected to enable robust integration of \ac{ML} components.

Towards this goal, we conducted a mixed-methods empirical study consisting of three stages.
First, we performed a \ac{SLR} to identify the challenges faced in (re-) architecting systems with \ac{ML} components and  the solutions proposed to meet them.  
We analysed \intRelevantArticles~relevant articles, from which we compiled an initial set of \intInitialChallenges~challenges and solutions. 
Second, we performed \intParticipants~semi-structured interviews with practitioners from \intCompanies~organisations~--~ranging from start-ups to large companies. 
The interviews were used to complement the initial set of challenges (and solutions) and  to assess the impact of each challenge on \ac{SA}.
In total, \intInterviewChallenges~new challenges were discovered in the interviews, as well as multiple new solutions.
Third, we ran a survey with \intSurveyParticipants~software architects in order to quantitatively validate and complement the challenges and solutions.
The survey also established a link between challenges, solutions and  software quality attributes, allowing the solutions to be restated as architectural tactics.

Overall, our main contributions are as follows.
First, we summarised academic and grey literature on the topic of \ac{SA} for \ac{ML} in a catalogue of \ac{SA} challenges and related solutions. 
This information can guide practitioners to (re-) architect software with 
\ac{ML} components, or as a gateway to relevant literature.
Second, we validated and complemented the initial findings by engaging with practitioners.
We found out that, although the initial challenges had solutions in the literature, the solution were considered incomplete by practitioners.
Third, we linked the architectural solutions to software quality attributes from the ISO/IEC 25010 standard~\cite{iso25010}, which allowed to restate them as architectural tactics.
Last, we assessed the impact of each challenge on \ac{SA}, which allowed us to contrast traditional 
\ac{SA} concerns with emergent \ac{ML} concerns.

The paper is organised as follows: first we discuss related work (Section~\ref{sec:background}) followed by the study design (Section~\ref{sec:study_design}), results (Section~\ref{sec:results}), discussions (Section~\ref{sec:discussion}) and conclusions (Section~\ref{sec:conclusions}).
All supplementary materials are public~\cite{anonymous_2021_5568797}.

\section{Background and Related Work}
\label{sec:background}

\ac{SE} for \ac{ML} is receiving increasing attention~\cite{nascimento2020software}. 
The related literature covers a broad range of topics; from \ac{SE} challenges raised by the adoption of \ac{ML} components~\cite{lwakatare2019taxonomy}, to practices~\cite{serban2020adoption}, guidelines~\cite{zhang2019software}, or design patterns~\cite{washizakimachine}.
Moreover, we consider the related field of \ac{SA} for big data and analytics platforms~\cite{sena2017characterizing}.
Therefore, we structure the presentation in three steps: first we introduce \ac{SE} challenges for \ac{ML} (with a focus on \ac{SA}), followed by solutions that meet the challenges and  by a discussion on \ac{SA} for big data and analytics, in the context of \ac{ML}.

Arpteg et al.~\cite{arpteg2018software} introduced twelve \ac{SE} challenges for \ac{ML}, classified in three categories: development, deployment and organisational.
From these, the challenges related to \ac{ML} platforms, to monitoring and logging \ac{ML} components and  to effort estimation for development and maintenance, were also identified in our \ac{SLR}.
Since Arpteg et al.~\cite{arpteg2018software} do not introduce solutions, the second and third stages of our study can be used to complement theirs.

Similarly, Ishikawa and Yoshioka~\cite{ishikawa2019engineers}, as well as Wan et al.~\cite{wan2019does}, studied how \ac{ML} impacts the traditional software development life-cycle.
Both studies are based on surveys and  have the bulk of responses from Asia. 
Notwithstanding this regional bias, they concluded that testing and evaluating the quality of \ac{ML} components is particularly difficult.
Distinct conclusions are drawn with respect to \ac{SA}. 
While Wan et al.~\cite{wan2019does} acknowledged \ac{SA} for \ac{ML} as difficult, Ishikawa and Yoshioka~\cite{ishikawa2019engineers} concluded that existing \ac{SA} methods apply equally to software with \ac{ML} components, although the tool support is immature.
We analysed the \ac{SA} challenges raised by \ac{ML} with finer granularity and  found out that while some challenges apply equally to software with or without \ac{ML} components,  \ac{ML} specific challenges (and solutions) also arise.

To classify the \ac{SE} challenges for \ac{ML}, Lwakatare et al.~\cite{lwakatare2019taxonomy} introduced a taxonomy, from which the challenges related to scalability and  serving were also identified in our study. 

An early publication that outlined \ac{SA} challenges and solutions for \ac{ML} was the work of~Sculley et al.~\cite{sculley2015hidden}.
The authors used the framework of technical debt to explore risk factors for \ac{ML} components.
Particularly, they argued that \ac{ML} components are subject to all maintenance issues specific to software components, as well as to new issues specific only to \ac{ML}.
Moreover, they introduced a set of anti-patterns and practices used to avoid technical debt.
Compared to Sculley et al.~\cite{sculley2015hidden}, the challenges (and solutions) introduced in this paper are broader and  consider more quality attributes. 

Breck et al.~\cite{breck2017ml} and Zhang et al.~\cite{zhang2020machine} studied the topic of testing for \ac{ML} components and  introduced testing and monitoring practices for different stages of the \ac{ML} development life-cycle.
While these practices are relevant to \ac{SE} for \ac{ML}, we are interested in the architectural decisions made for testing \ac{ML} components.
Therefore, we focus on higher-level decisions, such as using automating tests or designing the testing pipelines.

Amershi et al.~\cite{amershi2019software} conducted an internal study at Microsoft, aimed at collecting \ac{SE} challenges and practices for \ac{ML}.
They reported on a broad range of challenges and practices used at different stages of the \ac{ML} development life cycle.
In particular, modularity and component reuse in software with \ac{ML} components are challenges closely related to \ac{SA}, which are also discussed in this study.

Serban et al.~\cite{serban2020adoption}~and~Zhang et al.~\cite{zhang2019software} introduced two sets of \ac{SE} practices for \ac{ML} and deep learning, respectively.
While some practices are considered in \ac{SA}~--~\eg~the adoption of continuous integration~--~the broad selection of practices does not allow a focus on \ac{SA} (as in our study).
Therefore, the findings introduced can be used to complement theirs. 

Nascimento et al.~\cite{nascimento2020software} introduced a \ac{SLR} on the topic of \ac{SE} for \ac{ML} that analysed all articles up to 2019.
The authors observed that \ac{SA} is not yet a popular topic.
However, their taxonomy classifies software quality and infrastructure concerns separately from architecture.
We argue that such concerns are discussed extensively during \ac{SA} and  should be considered together when evaluating the popularity.

Washizaki et al.~\cite{washizaki2019studying} studied \ac{SA} patterns and anti-patterns for \ac{ML}, extracted from white and grey literature. 
Their proposal was followed by a larger study, where the initial set of patterns and anti-patterns was extended~\cite{washizakimachine}.
Their work is close to the first stage of our study (\ac{SLR}), where we identified a set of challenges and solutions in \ac{SA} for \ac{ML}.
We build upon it by enlarging the number of challenges and solutions and  by extensively validating our findings in the second and third stages of the study.
Moreover, although the challenges presented by~Washizaki et al.~\cite{washizaki2019studying} are recurrent, we found out that the solutions are not.
Therefore, we are cautious in using the taxonomy of design patterns.
Instead, we focus on smaller building blocks called tactics;
which bridge architecture decisions with quality attributes and  form the basis of design patterns~\cite{bass2003software, harrison2010architecture}. 

\begin{table*}[t]
    \centering
    \caption{Research questions for the \ac{SLR}.}
    \label{tbl:rqs_slr}
    \begin{tabular}{ L{0.5cm} L{8cm} L{8cm}}
    \toprule
    \textbf{ID} & \textbf{Research Question} & \textbf{Motivation} \\
    \midrule
    RQ1 & Which are the challenges reported in (re-) architecting software systems with \ac{ML} components? &  Understand the technical and organisational challenges, but also the requirements posed by adoption of \ac{ML} components.\\
    
    \midrule
    
    RQ2 & What solutions, tactics or patterns have been reported to successfully meet these challenges? &  Understand and identify solutions, tactics, or patterns for \ac{SA} with \ac{ML} components.  \\
    \bottomrule
    \end{tabular}
\end{table*}

The challenges raised by big data systems regarding continuous expansion of data volumes and the adoption of new technologies have been well studied and  several reference architectures have been proposed~--~\eg~\cite{avci2020software,sena2017characterizing,sena2018investigating}.
However, the proposals emphasise the data aspect of \ac{SA}, \ie~how to collect and manage various sources of data and satisfy performance demands of analytics platforms.
Therefore, although data visualisation and \ac{ML} components are present in the reference architectures, these do not record decisions taken for development, integration or serving \ac{ML} components.
Here, we focus on the latter, where the data aspect plays an important role but it is not 
the sole decision driver.
\section{Study design}
\label{sec:study_design}

Our study was organised in 3 stages and  consisted of a mixed-methods approach with a sequential exploratory strategy~\cite{easterbrook2008selecting}. 
In the first stage, we ran a \ac{SLR} to identify the challenges faced when (re-) architecting systems with \ac{ML} components and  the solutions proposed to meet them.
The second stage of the study consisted of semi-structured interviews, meant to complement and partially validate the data extracted in the first stage.
In the third stage, we ran a survey to gather quantitative data, augment and generalise the findings from the first two stages.
Data triangulation from multiple sources is known to increase the reliability of the results~\cite{easterbrook2008selecting}.
The design of the three stages is described below.

\emph{\textbf{Systematic Literature Review.}}
\Acp{SLR} are widely used in empirical \ac{SE} research and  provide a structured process to identify, evaluate and  interpret the information available regarding a research topic~\cite{kitchenham2007guidelines,easterbrook2008selecting, kitchenham2012systematic}.
\Acp{SLR} consist of three parts, namely defining a research protocol, conducting a review and reporting the results.
We followed the guidelines from~Kitchenham and Charters~\cite{kitchenham2007guidelines} and  defined a research protocol as follows.

\noindent
\textbf{Research questions.}
We aimed to gather evidence about the challenges faced when (re-) architecting software systems with \ac{ML} components.
Moreover, we looked for solutions that meet the challenges and synthesise practices, tactics or patterns.
Towards this goal, we formulated a set of research questions, summarised with their motivation in Table~\ref{tbl:rqs_slr}.
The questions facilitated the identification of challenges in the area of \ac{SA} with \ac{ML} components and  enabled the creation of an initial body of knowledge with solutions.

\begin{table}[t]
    \caption{Article selection for each information source.}
    \label{tbl:doc_selection}
    \centering
    \begin{tabular}{lcccc}
        \toprule
        \textbf{Source} & \textbf{\specialcell{Retrieved}} &  \textbf{\specialcell{Automatic \\ Filtering}} & \textbf{\specialcell{Manual \\ Inspection}} & \textbf{Used} \\ 
        \midrule
        ACM DL & \numprint{1000} & 647 & 21 & 7 \\ 
        IEEE Xplore & \numprint{1000} & 521 & 22 & 9 \\ 
        ScienceDirect & \numprint{1000} & 513 & 2 & 0 \\ 
        Scopus &\numprint{1000} & \numprint{732} & 2 & 0 \\ 
        \midrule
        Google Scholar & 100 & 100 & 7 & 3 \\ 
        Google & 100 & 100 & 12 & 10 \\ 
        \midrule
        Snowball & - & - & 16 & 13 \\ 
        \midrule
        Total & \numprint{4200} & \numprint{2613} & 82 & 42 \\ 
        \bottomrule
    \end{tabular}
\end{table}

%

\noindent
\textbf{Search strategy.}
To get a broad set of studies, we used multiple information sources.
First, we used automatic queries to retrieve studies from several digital libraries, namely IEEE Xplore, ACM Digital Library (ACM DL), Scopus and  ScienceDirect.
Shahin et al.~\cite{shahin2017continuous} observed that SpringerLink uses a different query mechanism than the others and  that Scopus indexes most articles from SpringerLink.
Therefore, in order to avoid inconsistencies in data retrieval, we relied on  Scopus.
Second, motivated by the findings of~Serban et al.~\cite{serban2020adoption}~--~which noticed that most literature on the topic of \ac{SE} for \ac{ML} consists of so called grey literature~--~we performed manual search in Google and Google Scholar, where the first 5 pages of results were inspected.
Last, we complemented the data set through a snowball strategy following references of relevant articles~\cite{budgen2008using}.

To define the search query, we followed the guidelines from~\cite{kitchenham2007guidelines} and  composed a string with synonyms of the words ``software architecture", ``machine learning", ``challenges" and  ``solutions".
After piloting several queries to validate the inclusion of previously known articles, we decided to use two distinct queries. 
The first query retrieved challenges in \ac{SA} for \ac{ML} and  the second query retrieved solutions.
The search string for the first query was:
``((``software architecture" OR ``software engineering" OR ``systems engineering") 
AND 
(``machine learning" OR ``deep learning" OR ``artificial intelligence" OR ``AI") \emph{AND (``challenge" OR ``problem" OR ``issue")})", where the emphasised string was replaced in the second query with:
\emph{AND (``solution" OR ``practice" OR ``guideline" OR ``tactic" OR ``pattern" OR ``architecture pattern" OR ``design pattern")}.
Using the word ``software" next to architecture or engineering helped to avoid articles from the general field of engineering (\eg~electrical engineering) or architecture.
Moreover, we observed that including both ``pattern" and ``design pattern" makes the query more effective.

\noindent
\textbf{Exclusion and inclusion criteria.}
Since the initial queries returned over \numprint{10000} results, we limited the answers to the first 500 articles for each data source and query. 
This reduced the number of articles to \numprint{1000} per source, which corresponds to recommendations and previous studies~\cite{maplesden2015performance, shahin2017continuous}.
Washizaki et al.~\cite{washizaki2019studying} and Nascimento et al.~\cite{nascimento2020software} showed that the majority of articles on the topic of \ac{SE} for \ac{ML} were  published after 2016.
Therefore, we also restricted our search to articles published after 2016.
Next, we \emph{automatically} filtered for duplicates and for records that contained the words ``proceedings" or ``workshop" in the title.
Moreover, we \emph{manually} excluded all opinionated articles; coming from companies or authors which could be traced back to companies that provide tools or services for \ac{SA}/\ac{SE} for \ac{ML}.
Thus, some bias regarding solutions driven by tools was avoided.
Since \ac{ML} for \ac{SE} receives increasing interest, we curated and removed the articles on this topic because our study focuses on \ac{SE} for \ac{ML}.
Also, we removed tool demonstration articles and  those not written in English.
In the final selection, we included all studies or grey literature articles that presented challenges or solutions based either on empirical studies or on experience (\eg~studies with empirical validation or organisation blogs describing their processes).

\begin{table*}[t]
    \centering
    \caption{Profiles of interview participants.}
    \label{tbl:interviees}
    
    \begin{tabular}{lp{11em}p{4em}p{3.5em}p{12em}p{5em}p{5em}p{3em}p{4.5em}}
        \toprule
        
         \textbf{ID} & \textbf{Position} & \textbf{Experience} & \textbf{Research} & \textbf{Org. Profile} & \textbf{Org. Size} & \textbf{Org. \ac{ML} Experience} & \textbf{Team Size} & \textbf{Deployment Interval} \\
         \midrule
        
         P1 & Solutions Architect & 3-5 years & No &  Technology (Internet) & \numprint{10000}+ & 6-9 years & 6-9 & 0-1 week \\ 
         
         P2 & System Architect & 3-5 years & Yes & Non-Technology (Transport) & \numprint{10000}+ & 3-5 years & 6-9 & 1-2 weeks \\ 
         
         P3 & Software Architect & 6-9 years & Yes & Technology (Automation) & \numprint{10000}+ & 3-5 years & 10-15 & 3-4 weeks \\ 
         
         P4 & Technology Lead & 3-5 years & Yes &  Technology (AI/ML) & 0-50 & 0-2 years & 10-15 & 0-1 week \\ 
         
         P5 & Software Architect & 6-9 years & Yes & Non-Technology (Medical) & 10 000+ & 3-5 years & 6-9 & 3-4 weeks \\ 
         
         P6 & Software Architect & 3-5 years & No & Technology (Automation) & 1000-5000 & 3-5 years & 10-15 & 1-2 weeks \\
         
         P7 & Director of Engineering & 6-9 years &  No & Technology (AI/ML) & 51-200 & 3-5 years & 10-15 & 1-2 weeks \\
         
         P8 & Senior Solutions Architect & 3-5 years & No & Technology (AI/ML) & 51-200 & 3-5 years & 6-9 & 3-4 weeks \\
         
         P9 & Head of Engineering & 3-5 years & No & Technology (Space)    & 51-200 & 0-2 years & 10-15 &  1-2 weeks \\ 
         
         P10 & CTO & 0-2 years & Yes & Technology (Robotics) & 0-50 & 1-2 years & 16-20 & 3-4 weeks \\

         \bottomrule
    \end{tabular}
    
\end{table*}

\noindent
\textbf{Study selection.}
After retrieving the initial set of~\allDocs~documents, we applied the selection criteria as follows.
In the first phase we applied the automatic filters, which reduced the set to~\filteredDocs~articles.
For these articles, we manually inspected the titles and the keywords and  selected~\manualWithoutSnowbal~articles to be completely assessed.
These were read completely and critically analysed, which reduced their number to~\intBeforeSnowball~relevant articles.
From their references,~\snowballAll~new articles were read, from which~~\snowball~were used in the final selection.
The distribution of articles and their sources for each stage of the review is presented in Table~\ref{tbl:doc_selection}.
We observe that although ACM DL and IEEE Xplore retrieved the bulk of articles for complete assessment, the grey literature search and snowballing strategies were more effective for the final selection.
Moreover, we observe that the distribution of articles per date 
resembles the one from~\cite{washizaki2019studying,nascimento2020software}~--~\ie~the number of articles is increasing year by year. 
Similarly, by analysing the distribution of articles based on the venue type 
we note that the majority of academic articles were published in conferences or workshops and not in journals.
We conjecture that: (i) \ac{SA} for \ac{ML} is an emerging field and  the publications did not reached the maturity needed for journal publication and  (ii) journals have a longer publication cycle, therefore many publications may be in review. 
The complete list of articles, their sources and a demographic characterisation is attached in the supplementary materials (Appendix A and B).

\noindent
\textbf{Data extraction and synthesis.}
From all articles, we extracted and classified the information in: (i) demographics and context, (ii) \ac{SA} for \ac{ML} challenges (RQ1), (iii) \ac{SA} for \ac{ML} solutions and tactics (RQ2),
(iv) data types used.
To analyse the demographics data we used descriptive statistics.
To extract the data for (ii) and (iii), we used qualitative analysis methods. 
In particular, we used thematic analysis~\cite{braun2006using}, which defines a process based on the following 5 steps: (1) familiarity with data~--~the articles were examined to form initial ideas, (2) initial code generation~--~the initial list of challenges and solutions was extracted, (3) theme search~--~common elements between the challenges and the solutions, respectively, were identified,
(4) theme review~--~challenges and solutions were compared and  common items were merged or dropped, (5) definition and naming~--~each challenge and solution was defined and named.

\emph{\textbf{Interviews.}}
To complement the results from the \ac{SLR}, we conducted 10~interviews with practitioners.

\noindent
\textbf{Protocol.} The interview protocol was designed following the guidelines from~Hove and Anda~\cite{hove2005experiences} and  consisted of~\intQuestions~questions designed to support a natural conversation between the participants.
All interviews were conducted online, through video calls (\intVideo~interviews) or e-mail (\intMail~interviews).
To enable participants to become familiar with the interview objectives~\cite{hove2005experiences}, we shared a shorter version of the interview plan at least three days before the meeting.

The interviews were structured in~\intSections~sections.
First, we described the research goals and background.
Second, we asked participants to share information about their background and demographics.
Third, we asked to participants describe the constraints and challenges faced in 
\ac{SA} for \ac{ML} either for the last project they worked on or for a specific project.
This part enabled a discussion about the challenges faced (and the solutions adopted) and  was meant to complement the data obtained previously.
Next, we asked participants to comment on each of the challenges from the \ac{SLR}, evaluate their impact on \ac{SA} and  propose solutions.
Last, we asked participants to provide open-ended comments.
We continuously refined the questions and 
after the first three interviews the questions remained stable. 
Two questions were merged due to redundancy and one was modified to be more descriptive.

\noindent
\textbf{Participants.} The interview participants were recruited using purposeful sampling~\cite{palinkas2015purposeful}. 
We contacted participants with experience in (re-) architecting systems with \ac{ML} components, or involved in architectural decisions (\eg~had the role of architect, or a leading position in engineering) and  who are working (or worked) for companies using \ac{ML}.
To identify the participants, we used our personal network of contacts.
Moreover, we compiled a list of organisations that use \ac{ML} from outlets such as Forbes or MIT Technology Review.
Later, we traced back candidates from the organisations (holding positions linked to \ac{SA}) through LinkedIn and  contacted them.
The list of interview participants, together with the demographics data regarding their background and team characteristics are presented in Table~\ref{tbl:interviees}.
We observe that the participants' background is diverse, ranging from software and system architects to engineering leaders and CTOs.
Moreover, the participants' and organisations' experience is diverse~--~ranging from start-ups to large organisations with vast experience in \ac{ML}. 
Since the organisations had different profiles, we classified them into (i) \emph{Technology}~--~focus on developing technology products and  (ii) \emph{Non-Technology}~--~do not focus on technology products, but use \ac{ML} for their processes.
We also note that many participants had previous experience in research, being directly involved or in close collaborations with research groups.
We hypothesise that the research driven process for \ac{ML} is a contributing factor to this result.

\noindent
\textbf{Data analysis.} 
The interviews were processed using thematic analysis, a technique which consists of the five steps recommended by~Cruzes and Dyba~\cite{cruzes2011recommended}: (i) data extraction~--~the interviews were transcribed, read and  key points were extracted, (ii) data coding~--~the initial \ac{SA} challenges and tactics, as well as the impact of each challenge on \ac{SA} (\eg~low or high impact) were defined, (iii) code to themes translation~--~for each transcript the initial codes were combined into potential themes (\eg~automated testing), (iv) high-order theme modelling~--~the themes were compared and merged, or dropped if the evidence was not sufficient (\eg~automated testing was merged in CI), (v) synthesis assessment~--~arguments for the extracted data were established, for example in terms of credibility (if the core themes were supported by the evidence) or confirmability (if there was consensus among the authors on the coded data).

\emph{\textbf{Survey.}}
To generalise the findings with a large sample size and augment the solutions, we ran an online survey.
The survey was developed using the guidelines from~Kitchenham and Pfleeger~\cite{kitchenham2008personal} 
and~Ciolkowski et al.~\cite{ciolkowski2003practical}.
We designed a cross-sectional observational study asking participants at the moment of taking the survey which solutions they adopt, for each challenge.
Moreover, we asked participants about their background in order to assign them to groups; making the study a concurrent control study in which participants are not randomly assigned to groups.

\noindent
\textbf{Questionnaire.} 
The questionnaire consisted of five sections. 
In (i) the preliminaries we asked participants about their background (5 questions), to select a recent project where they played a role in \ac{SA} for \ac{ML} and  to provide information regarding the challenges faced, the project constraints and  the data types used (3 questions).
Next, we asked participants to (ii) select or propose new solutions for the challenges identified previously (20 questions).
Since multiple solutions involved instrumentation, monitoring or alerts, we added a question regarding the architectural decisions for designing these modules (1 question).
Afterwards, we asked participants to (iii) select the architectural style (if any) adopted in  their project (1 question) and  to (iv) link the solutions to software quality attributes (1 question).
The questionnaire ended with a section where (v) participants could provide open ended feedback (1 question).

The answers allowed multiple choices, with the solutions extracted from the \ac{SLR} and from the interviews.
Besides, we provided an open answer called 'Others', where participants could propose new solutions.
The quality attributes used in the fourth section
were extracted from the ISO/IEC 25010 standard~\cite{iso25010}, which is widely regarded as mature. 
However, we found the ``Installability" and ``Replaceabiliy" attributes out-dated and  replaced them with ``Deployability"; which better reflects deployment and roll-back.

\noindent
\textbf{Survey Pilot.}
Before distributing the survey, we invited four candidates to assess the survey in our presence and  suggest improvements.
The participants did not consider any question redundant.
Using their feedback, we added three new answers possible to the questions where the answers were considered incomplete and rephrased other answers and questions. 

\noindent
\textbf{Distribution.}
To distribute the survey, we used a snowballing strategy.
At first, we reached out to our network of contacts, asked them to fill in the survey and forward it to potential candidates.
Second, we expanded the list of contacts from interview recruitment.
In total, we sent \intMails~e-mails or private messages to potential participants.
Third, we advertised the survey through open channels used by practitioners, \ie~Reddit and LinkedIn.

\noindent
\textbf{Data Analysis.}
We processed the standard answers using descriptive statistics and  the open-ended answers using thematic analysis.
Moreover, we analysed the association between the adoption of solutions using the Phi coefficient.
\section{Results}
\label{sec:results}
We present the results from the three stages of the study as follows: (i) the \ac{SLR} results in Section~\ref{sec:slr_results}, (ii) the interview results in Section~\ref{sec:interviews} and  (iii) the survey results in Section~\ref{sec:survey}.
 
\subsection{Results from the SLR}
\label{sec:slr_results}

\begin{table*}[ht]
	\caption{List of \ac{SA} challenges for \ac{ML}, and related solutions.}
	
	\label{tbl:practices}
	\begin{tabular}{p{1em}p{2.7em}p{21em}p{26.5em}p{6em}}
    \toprule
		\textbf{Nr.} & \textbf{Category} & \textbf{Challenges} & \textbf{Solutions}  & \textbf{References}  \\
    \midrule
        1 & Reqs. & At design time the information available is insufficient to understand the customers or the projects. &
        Run simulations to gather data. Use past experience. Measure and document  uncertainty sources.
        & \cite{ishikawa2019engineers,belani2019requirements,chechik2019uncertain, lwakatare2019taxonomy,liu2020emerging} \\ 

        2 & Reqs. & \ac{ML} components lack functional requirements. & 
        Use metrics as functional requirements. Include understandability and explainability of the outputs.
        & \cite{ishikawa2019engineers,belani2019requirements,lwakatare2019taxonomy,de2019understanding,chechik2019uncertain} \\

        3 & Reqs. & \ac{ML} projects have regulatory restrictions and may be subject to audits. & 
        Analyse regulatory constraints up-front. Adopt an AI code of conduct. Design audit trails.
        & \cite{kaur2020requirements,national2019national,MLRARCH,serban2020adoption} \\
        
        \midrule
        
        4 & Data & Data preparation may result in a jungle of scrapes, joins, and sampling steps, often with intermediate outputs. & 
        Design separate modules/services for data collection and data preparation. Integrate external tools. 
        & \cite{sculley2015hidden,li2017scaling,de2019understanding} \\

        5 & Data & Data quality is hard to test, and may have unexpected consequences. & 
        Design separate modules/services for data quality assessment. Integrate external tools.
        & \cite{zhang2019software, polyzotis2019data,li2017scaling,mcgraw2019security} \\
        
        \midrule
        
        6 & Design & Separate concerns between training, testing, and serving, but reuse code between them. & 
        Standardise model interfaces. Use one middleware. Reuse virtualisation, infrastructure and test scripts.
        & \cite{amershi2019software,RLML,wu2019machine} \\ 
        
        7 & Design & Distinguish failures between \ac{ML} components and other business logic. &
        Separate business logic from ML components. Standardise interfaces and use one middleware between them.
        & \cite{yokoyama2019machine,rahman2019machine} \\           
        
        8 & Design & ML components are highly coupled, and errors can have cascading effects. 
        & 
        Design independent modules/services for \ac{ML} and data. Standardise interfaces and use one middleware. Relax coupling heuristics between \ac{ML}  and data. 
        & \cite{wan2019does,northrop2019designing,horneman2020ai} \\

        9 & Design &  \ac{ML} components bring inherent uncertainty to a system. & 
        Use n-versioning. Design and monitor uncertainty metrics. Employ interpretable models/human intervention.
        & \cite{aniculaesei2018toward,northrop2019designing,serban2019designing,serban2020towards,horneman2020ai} \\
        
        10 & Design & ML components can fail silently. 
        These failures can be hard to detect, isolate and solve. 
        & 
        Use metric monitoring and alerts to detect failures. Use n-versioning. Employ interpretable models.
        & \cite{biondi2019safe, serban2019designing, woods2016software} \\
        
        11 & Design & \ac{ML} components are intrinsically opaque, and deductive reasoning from the architecture artefacts, code or metadata is not effective. 
        & Instrument the system to the fullest extent. Use n-versioning. Employ interpretable models. Design log modules to aggregate/visualise metrics.
        & \cite{scheerer2020towards,northrop2019designing,horneman2020ai,RLML} \\

        12 & Design & Avoid unstructured components which link frameworks or APIs (\eg~glue code). & Wrap components in APIs/modules/services. Use standard interfaces and one middleware. Use virtualisation. & 
        \cite{sculley2015hidden} \\

        13 & Design & Automation and understanding of \ac{ML} tasks is difficult (AutoML).
         & Version configuration files. Design the log and versioning systems to support AutoML data retrieval.
         & \cite{wan2019does,li2017scaling,wu2019machine,serban2020adoption,PAPML} \\        
        
        \midrule

        14 & Testing & \ac{ML} testing goes beyond programming bugs to issues that arise from model, data errors, or uncertainty. &  
        Design model and data tests. Use CI/CD. Use integration and unit tests. Use data ownership for test modules.
        & 
        \cite{arpteg2018software,amershi2019software,nishi2018test,zhang2019software,reimann2020achieving}  \\
        
        15 & Testing & Validation of \ac{ML} components for production is difficult. & 
        Use metrics and CI/CD for validation.  Use alerts, visualisations, human intervention. Design release processes. 
        &  \cite{CD4ML}  \\

        \midrule
        
        16 & Ops. & ML components require continuous maintenance, retraining and evolution. & 
        Design for automatic continuous retraining. Use CI/CD. Use automatic rollback. Use infrastructure-as-code. Adopt standard release processes.
        & \cite{CD4ML,CTPML,northrop2019designing,washizaki2019studying,wan2019does,zhang2019software,liu2020emerging} \\

        17 & Ops. & Manage the dependencies and consumers of ML applications. & 
         Encapsulate \ac{ML} components in identifiable modules/services. Use authentication and access control. Log consumers of \ac{ML} components.
         & \cite{sculley2015hidden,DACH,yokoyama2019machine,batyuk2018software,horneman2020ai} \\         
        
        18 & Ops. & Balance latency, throughput, and fault-tolerance, needed for training and serving. & 
        Design for batch processing (training) and stream processing (serving), \ie~lambda architecture. Physically isolate the workloads. Use virtualisation.
        & \cite{li2017scaling,DMLA,washizakimachine, wu2019machine,ADPML} \\

        19 & Ops. & Trace back decisions to models, data and reproduce past results. & 
        Design for traceability and reproducibility; log pointers to versioned artefacts, version configurations, models and data.
        & P10   \\

        \midrule
        
        20 & Org. & \ac{ML} applications use heterogeneous technology stacks which require diverse backgrounds and skills.  & 
        Form multi-disciplinary teams. Adopt an AI code of conduct. Define processes for decision-making. Raise awareness about \ac{ML} risks within the team.
        & P1 \\

    \bottomrule 
    \end{tabular}
\end{table*}

From the \ac{SLR} we identified an initial set of \intInitialChallenges \ challenges, presented in Table~\ref{tbl:practices}.
We note that the \ac{SLR} data have numerical references.
To classify the practices, we used a custom taxonomy because the \ac{ML} taxonomy is different than the traditional \ac{SE} taxonomy~\cite{amershi2019software}.
Moreover, \ac{ML} taxonomies are divergent~\cite{serban2020adoption}.
For example,~Amershi et al.~\cite{amershi2019software} present a nine-stage taxonomy for the \ac{ML} process, while Sato et al.~\cite{CD4ML} use only six stages. 
These taxonomies have roots in the CRISP-DM model~\cite{wirth2000crisp}. 
However, recent studies show these models are not fit for all contexts~\cite{haakman2020ai}.
Since existing taxonomies are divergent, we constructed a broad taxonomy compatible with previous work and focusing on~\ac{SA}.

The taxonomy was used to classify the challenges (and solutions) in:
(i) \emph{Requirements (Reqs.)}~--~requirements elicitation for \ac{ML} components, mapped to model requirements and business understanding~\cite{amershi2019software,wirth2000crisp}, (ii) \emph{Data}~--~data collection, preparation and validation, mapped to data taxonomies~\cite{amershi2019software,serban2020adoption,wirth2000crisp}, (iii) \emph{Design}~--~the system's structure, \ac{SA} decisions and trade-offs, mapped to training and coding taxonomies~\cite{amershi2019software,serban2020adoption}, (iv) \emph{Testing}~--~testing and validation of software with \ac{ML} components, mapped on the evaluation taxonomies~\cite{amershi2019software,wirth2000crisp} and  (v) \emph{Operational (Ops.)}~--~deployment, monitoring and evolution, mapped to deployment taxonomies~\cite{amershi2019software,wirth2000crisp}.

\emph{\textbf{RQ1.}} Answering RQ1 from Table~\ref{tbl:rqs_slr}, we identified \intInitialChallenges~challenges through the \ac{SLR}, classified in five categories.
The \emph{Reqs.} challenges focus on the inability to understand a project and estimate the effort upfront. 
Moreover, the opaque nature of \ac{ML} components~--~for which functional requirements are difficult to define and  which may be subject to regulatory restrictions, emerged as challenging.

\begin{figure}[t]
    \centering
    \includegraphics[width=8.5cm, keepaspectratio]{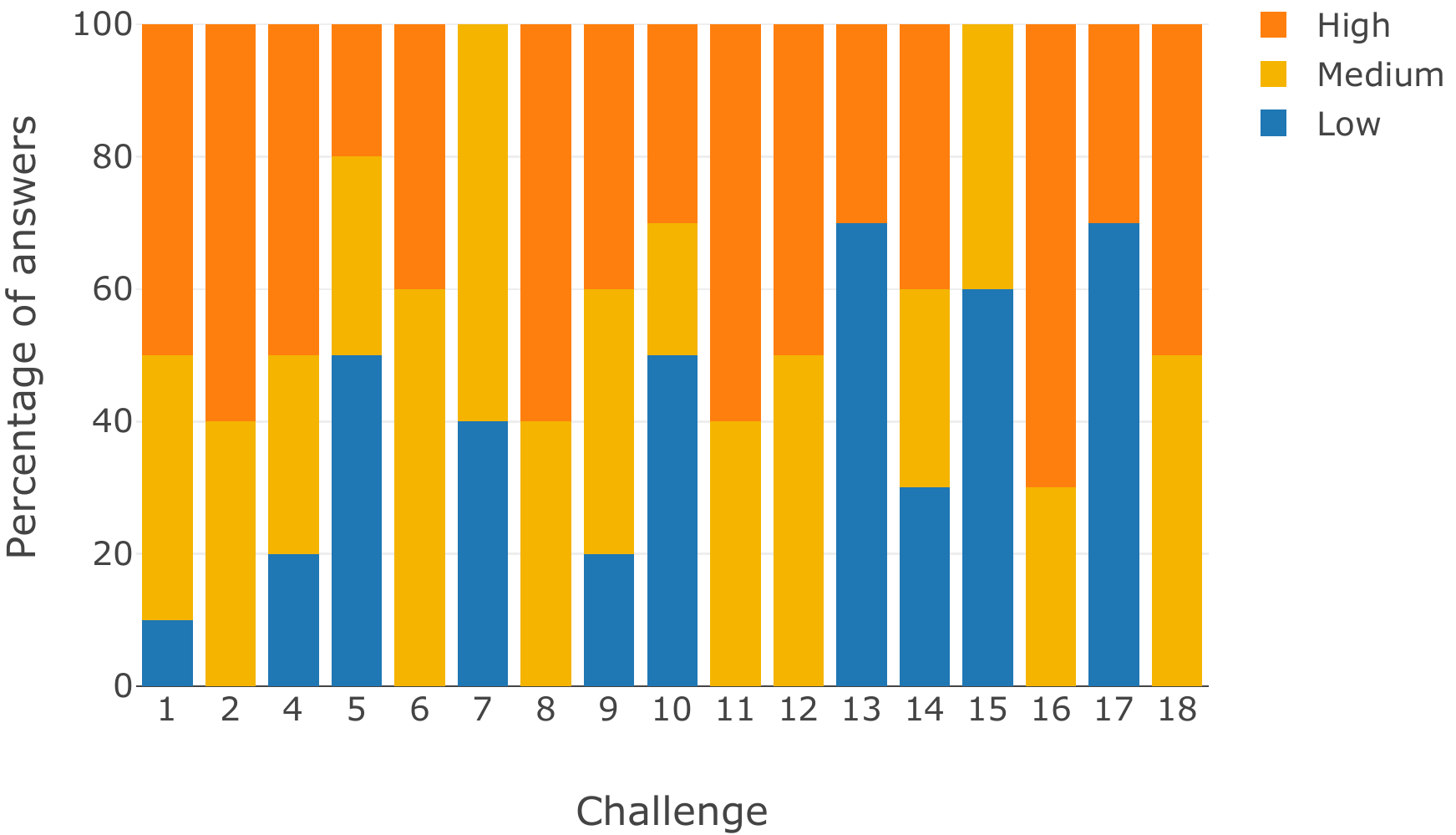}
    \caption{The impact of the challenges from Table~\ref{tbl:practices} on \ac{SA}, as assessed by interview participants.}    
    \label{fig:impact_interviews}
\end{figure}

The \emph{Data} challenges relate to data preparation and data quality assessment.
This result contrasts previous concerns from big data and analytics platforms~\cite{sena2017characterizing}, where the focus was on data storage and accessibility.
Nonetheless, this result corresponds with the expectation that \ac{ML} components are evolved from big data platforms and  therefore extend and overcome the challenges met there.

The largest category of challenges, \emph{Design}, includes both traditional \ac{SA} challenges, such as managing component coupling and  new \ac{ML} specific challenges, such as managing inherent uncertainty or designing for development automation (AutoML).
We also notice a challenge regarding the integration of \ac{ML} components with traditional software components and business logic (7), which finds it difficult to distinguish failures between the two.

In contrast to Design, the \emph{Testing} challenges are \ac{ML} specific.
Here, the focus is on model testing~--~which goes beyond programming bugs~--~and on validation for production~--~which does not rely on new features or bug fixes, but on measurements that must meet multiple criteria. For example, accuracy, robustness or bias.

In the \emph{Ops.} category, the challenges relate to deployment, maintenance and  resource usage between training and testing.
We observe that maintenance of \ac{ML} components is based on retraining and deploying models trained with new data, which erodes the boundaries between maintenance and evolution.

\emph{\textbf{RQ2.}} Answering RQ2 from Table~\ref{tbl:rqs_slr}, through the \ac{SLR} we found distinct solutions to each challenge in Table~\ref{tbl:practices}. 
A detailed list of solutions extracted from the \ac{SLR} is presented in Appendix D
while Table~\ref{tbl:practices} presents the solutions from all stages of the study.
We note that \intPercSLR\% of solutions came from the \ac{SLR}, while the rest came from later stages of the study. 

\subsection{Results from the interviews}
\label{sec:interviews}

The interviews were meant to qualitatively assess and complement the \ac{SLR} data.
As mentioned in Section~\ref{sec:study_design}, the interviews had specific questions to discover new challenges, to evaluate the impact of each challenges on \ac{SA} and to propose new solutions. 

Two new challenges were added after the interviews and  several others were reinforced.
The first new challenge, (19), relates to tracing back serving decisions to \ac{ML} models and data and  to the ability to accurately reproduce past experiments.
This challenge brings together two concepts~--~traceability and reproducibility~--~both known to raise issues in \ac{ML}~\cite{pineau2020improving}.
Only one interview participant mentioned this challenge can have a significant impact on \ac{SA}.
Nonetheless, we included it, in spite of the fact that we did not have convincing evidence and  sought validation with the survey.

The second challenge, (20), relates to managing multi disciplinary teams, which use heterogeneous technology stacks (\eg~\ac{ML} frameworks, infrastructure scripts, business logic). 
Since this challenge does not fit any previous class, we defined a new class~--~\emph{Organisation (Org.)}~--~which gathers organisation wide concerns that fall in the attributes of software architects.
This class aligns with the view that software architects shall consult and bridge multiple teams, which solve problems beyond \ac{SA}~\cite{kruchten2008software}.
The challenge was mentioned by one participant, part of a large organisation with well established teams who work at different levels of the technology stack.
Therefore, the solution was to form multi-disciplinary teams which can work close together and  adopt standard ways of working.
No participant from small organisations raised this challenge, which begs the question if small organisations are more agile and  can overcome it.
The answer to this question was sought with the survey.

 \begin{figure*}[t]
 	\centering
 	\subfloat[Distribution of participants by demographics.\label{fig:sankey}]{\includegraphics[width=7.11cm, keepaspectratio,valign=t]{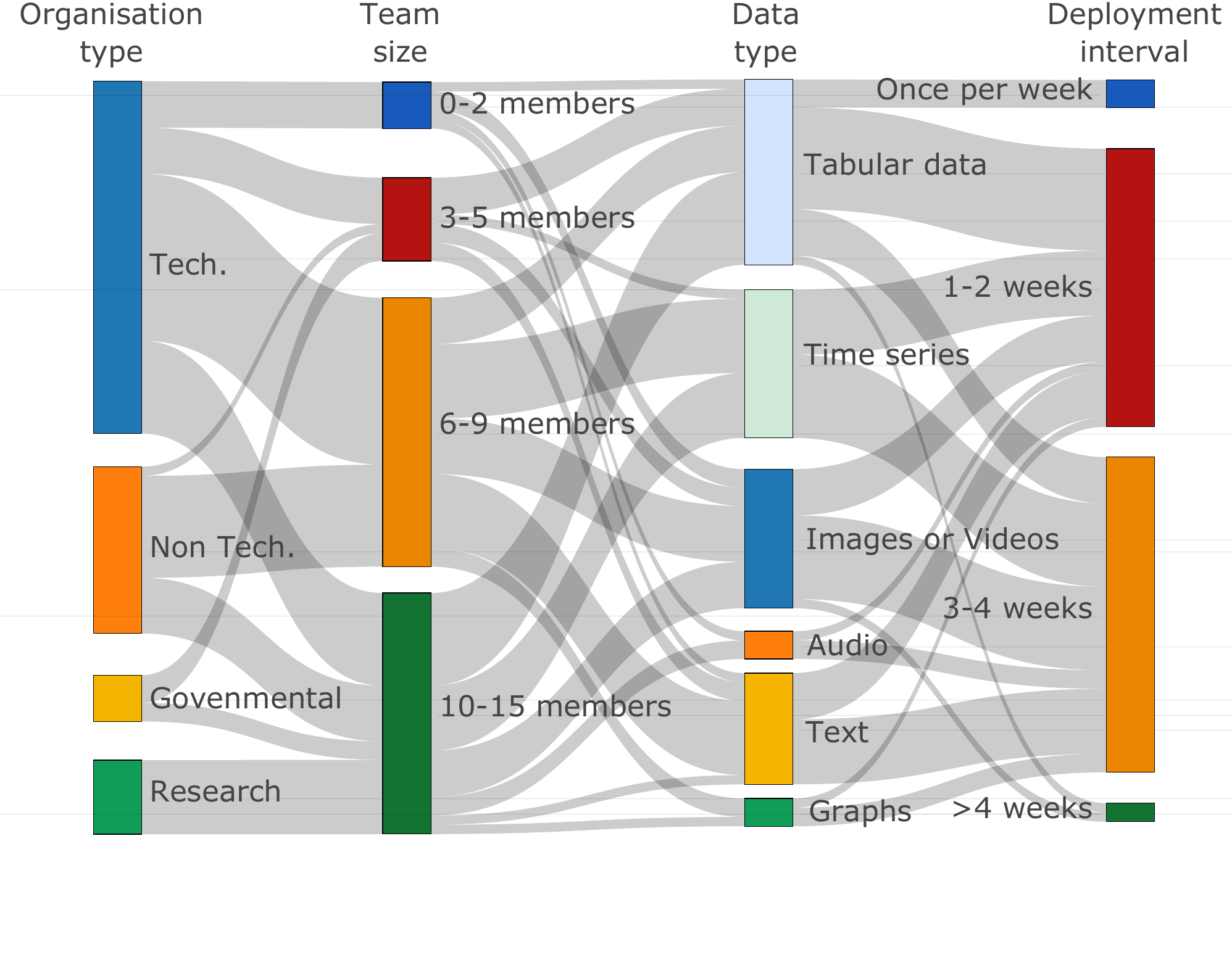}}
 	\subfloat[\ac{SA} decision drivers.\label{fig:survey_decision_drivers}]{\includegraphics[width=5.435cm, keepaspectratio,valign=t]{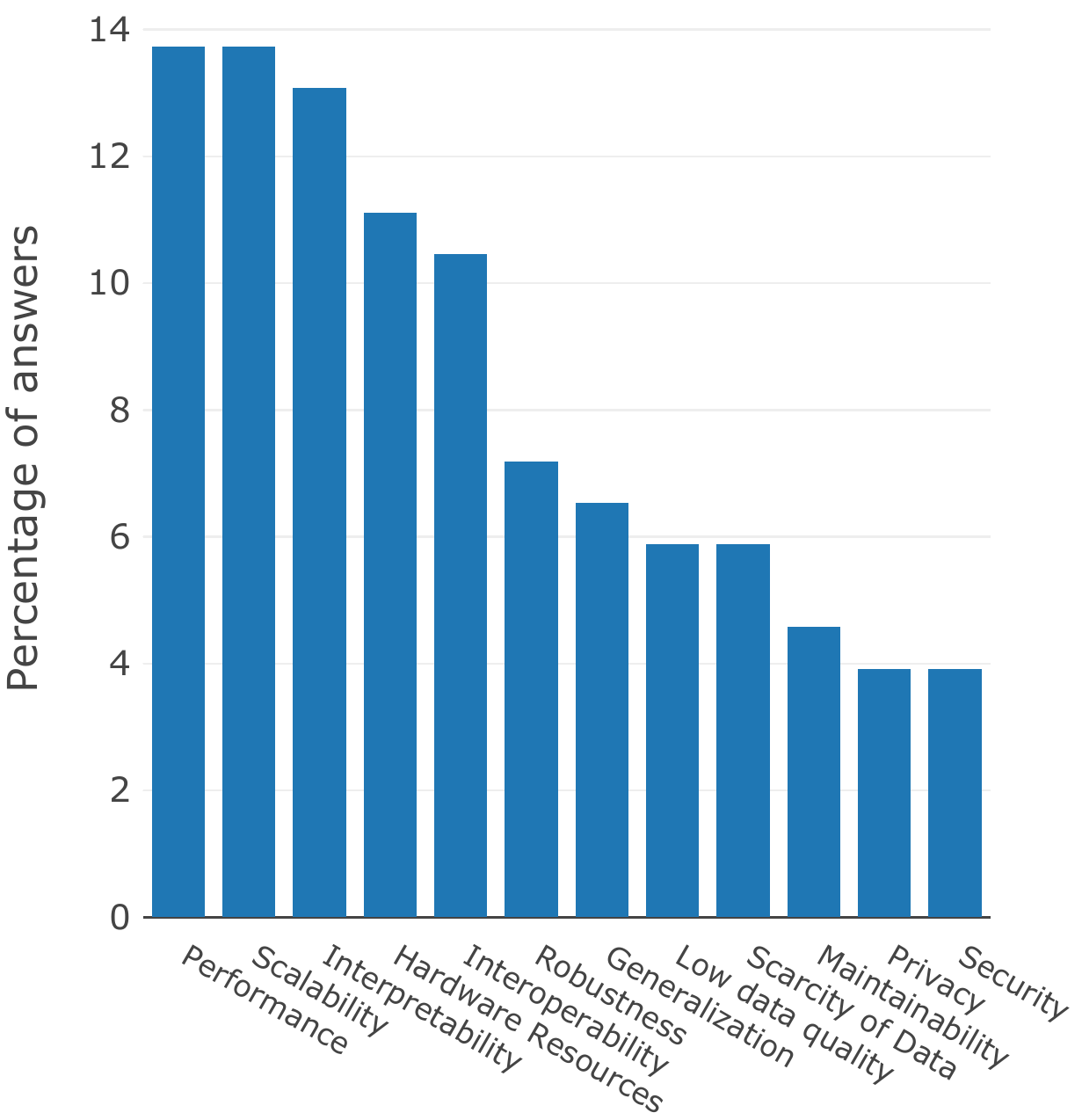}}
 	\subfloat[Architectural styles.\label{fig:survey_arch_tyles}]{\includegraphics[width=5.56cm, keepaspectratio,valign=t]{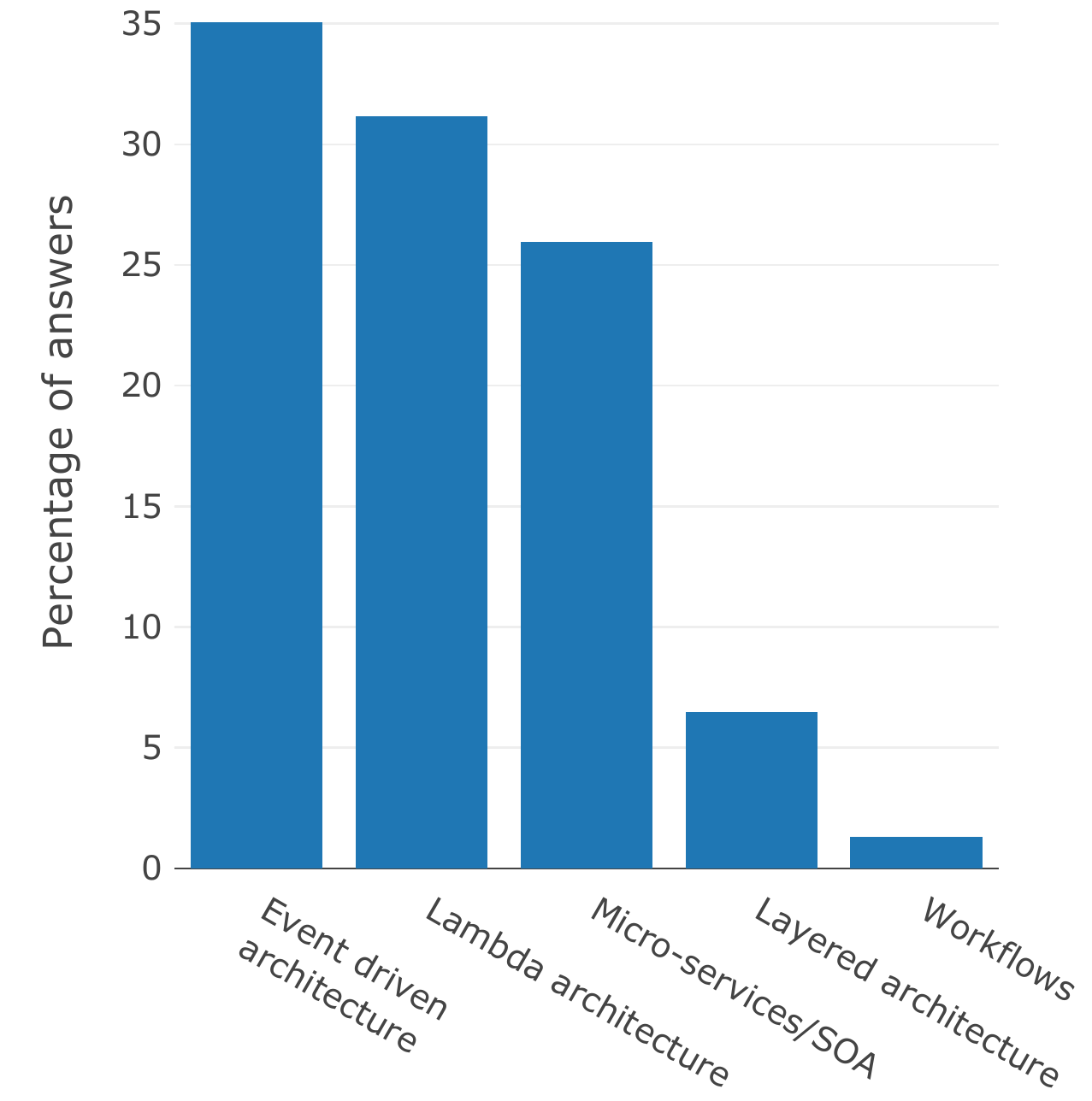}}
 	\caption{Characterisation of survey results using (a) demographics, (b) \ac{SA} decision drivers and  (c) architectural styles.} 	 	
  	\label{fig:survey} 
 \end{figure*}

We also asked participants about the most important architectural decision drivers and  about the data types used.
We note that ``Scalability", ``Hardware" constraints and  ``Data" concerns were mentioned as main decision drivers, followed by ``Interpretability".
Together with the data type used, we could also identify the main decision drivers for specific data types.
Here, we note that participants using Images \& Videos or Time Series found ``Scalability" and ``Hardware" constraints as the main decision driver.
Moreover, participants using Simulations were also driven by ``Hardware" constraints.
We also observed a new decision driver~--~called ``Generalisation"~--~which describes the ability of a \ac{ML} component to maintain training performance in production.
This driver is related to challenge (10) and  the solution suggested by participants was to use n-versioning; \ie~multiple versions of the \ac{ML} models (some of which may be more trustworthy).

While evaluating the challenges extracted from the \ac{SLR}, we asked participants to assess their impact on \ac{SA}.
The results are illustrated in Fig.~\ref{fig:impact_interviews} and  use an ordinal scale with three possible options: low, medium or high impact.
Challenge (3) could not be evaluated because the participants did not report regulatory restrictions.
We believe this result is due to the fact that \ac{ML} regulations are still in draft phase or not yet enforceable~\cite{jobin2019global}.
Within the challenges with the highest impact, we observe one traditional challenges that is strengthened by \ac{ML} (component coupling (8)) and  multiple \ac{ML} specific challenges.
For example, opaqueness of \ac{ML} components (11) or training-serving resource management (18).
The highest impact on \ac{SA} comes from the need to continuously retrain \ac{ML} components (16), while the lowest impact comes from  \ac{ML} task automation (13).

Besides challenges, the interviews allowed us to complement the initial set of solutions. 
In total, 48\%~new solutions came from the interviews.
Participants provided new solutions for all challenges besides challenges (2) and (3). 
While the solutions for (2) were regarded as complete, (3) was disregarded because participants did not reported regulatory constraints.

During the thematic analysis, we combined several solutions by bridging \ac{ML} and \ac{SE} terminology, while striving for conciseness. 
Here, we describe resulting themes which may be ambiguous due to name compression.
Using the same type of interfaces for business logic and \ac{ML} components, for all \ac{ML} components or within all projects was modelled as the use of ``standard interfaces". 
Participants reported multiple techniques to standardise the interfaces, \eg~REST APIs, gRPC or more general contracts for service oriented architectures.
While the techniques are project specific, the architectural decision to unify the interfaces is singular.

Moreover, using multiple versions of a \ac{ML} model~--~also called ensembles of models in \ac{ML}~--~is similar to n-version programming. 
Therefore, we grouped these solutions in the ``n-versioning" theme.

The separation of concerns and encapsulation of code was modelled as one theme: ``design separate modules/services".
Here, participants reported that code was either developed as separate modules
or as independent services.
This development included encapsulation for reuse.

Furthermore, we defined the use of the same middleware for business logic and \ac{ML} components in training and serving as ``use one middleware" and  the development of dashboards or modules to analyse the \ac{ML} components in ``visualisations".
Comprehensive materials and traces for the interview analysis can be found in Appendices E and F, while a detailed description of the themes in J.

\subsection{Results from the survey}
\label{sec:survey}

In total, we received \intSurveyAll~answers from which we filtered out (using the preliminary questions) respondents who did not play a role in \ac{SA} for \ac{ML}.
Moreover, we filtered out respondents who spent less than two minutes fulfilling the survey and  respondents who answered less than 50~\% of the preliminaries or less than 50\% of the technical questions. This process ensured only thoughtful answers were used in the analysis and  entailed \intSurveyParticipants~complete answers.

\emph{\textbf{Demographics.}}
We first grouped respondents by demographics.
We note that the majority of respondents (\intTechOrgPercentage\%) work for Technology organisations and  have between 3-5 (\intExpfiveyears\%) or 1-2 (\intExptwoyears\%) years of experience.
These results align with related work~\cite{ishikawa2019engineers,serban2020adoption} and  are in line with expectations that Technology companies are early adopters of \ac{ML} technologies.
Other groups are also represented, \ie~Non-Technology (\intNonTechOrgPercentage\%), Governmental Org. (\intGovOrgPercentage\%) and Research labs (\intResearchOrgPercentage\%).
Similarly, beginners which just started (\intExpbeg\%) and very experienced respondents, with 6-9 years of experience (\intExpnine~\%), are well represented. 

We also grouped respondents by regions into Europe (\intEurope\%), North America (\intUSA\%) and Asia (\intAsia\%).
Here, we observe a slight over-representation of Europe and  under-representation of Asia.
The possible bias stemming from the grouping by regions will be discussed in Section~\ref{sec:discussion}.

As with the interviews, we asked respondents about their team size, data types used and  deployment intervals.
This data is illustrated in Fig.~\ref{fig:sankey}, where the height of the bars represents the percentage of respondents and  the height of the connections represents the percentage of respondent who fall in the target class.
We observe that the majority of respondents belong to teams between 6-9 (\intTeamNine\%) or 10-15 (\intTeamFifteen\%) members.
In particular, the majority of Technology and Non-Technology teams have between 6-9 members, while the majority of Research teams are larger, between 10-15 members. 

Regarding data types, we observe that, with the exception of Audio and Graphs, the data types have similar distributions.
Moreover, the majority of respondents using Tabular data deploy new versions between 1-2 weeks, while respondents using Images, Videos, Audio or Text between 3-4 weeks.
We conjecture that this result relates to the \ac{ML} techniques suitable to each data type, \ie~Images, Videos, Audio or Text models are based on deep learning, require longer training times and the collection of larger data sets.
In contrast, Tabular data can be processed with more traditional \ac{ML} techniques (\eg~Random forest).
These techniques require smaller data sets and training time; making the teams more agile.

Overall, the demographics indicate that our survey data is diverse and  resembles data from interviews and related work~\cite{ishikawa2019engineers,serban2020adoption}.
More details about the demographic analysis can be found in Appendix H.

\emph{\textbf{Decision drivers.}}
Second, we asked respondents about the most important decision drivers in their projects.
This data is illustrated in Fig.~\ref{fig:survey_decision_drivers}.
We observe that ``Scalability" and ``Hardware" are consistent with the interviews and  occupy leading positions.
To better understand the data challenges, we divided them into "Low data quality" and "Scarcity of data".
Taken together, data related concerns are consistent with the interviews.
Separately, they were considered equally important, but none of them ranks high.

We also note that ``Performance", ``Interpretability" and  ``Interoperability"  rank higher for survey respondents, while ``Privacy" and ``Security" rank very low.
This result is cause of concern, since documents from policy makers and advisory bodies suggest these topics are paramount for trustworthy development of \ac{ML}~\cite{EUGuidelines}.
We conjecture that, although  a large body of academic literature on security of \ac{ML} exists, it is still limited in its applicability.
For example, all defences against adversarial examples~--~a known threat for \ac{ML} components~--~have been breached~\cite{carlini2019evaluating}.
The data also indicates that respondents prioritise operational quality attributes, such as scalability or performance in spite of security or privacy.

\emph{\textbf{Solutions to challenges in Table~\ref{tbl:practices}.}} 
Third, using the survey results we filtered out and ranked the solutions from previous stages of the study.
In particular, we considered the solutions that were selected less than 5\% of the time as not relevant and  filtered them out.
Moreover, we used the number of times the solutions were selected by respondents to rank them.
The ranking is reflected in Table~\ref{tbl:practices} by the order in which the solutions are presented.

For all challenges, respondents could also suggest new solutions or provide comments using the "Other" field.
In total, we received three suggestions and open comments.
We analysed the results using thematic analysis and  found out that all suggestions were variations of the solutions provided or comments suggesting some solutions do not apply.
For example, one respondent mentioned that, due to tight performance constraints, it was not possible to apply n-versioning.
The comment suggests that, given exceptional constraints, some solutions do not apply.
This result is expected since the first two stages of the study strove for generalisation and outliers may exist.
Nonetheless, the lack of novel suggestions for solutions brings evidence that the first two stages of the study entailed comprehensive solutions to all challenges.

We also note that some solutions are recurrent and can be applied to multiple challenges.
For example, the use of standard interfaces for \ac{ML} components and business logic or the use of interpretable models.
These results are expected since architectural decisions may impact multiple aspects.

As mentioned previously, an extra question was added for decisions regarding instrumentation, monitoring and alerts.
The solutions were inspired by interviews; where participants reported the development of independent logging, alert or visualisation modules.
Here, we note that the majority of survey respondents reported the development of independent modules/services for instrumentation and monitoring.
Moreover, respondents reported on separating logging concerns between training and serving and  on the development of independent modules to aggregate and visualise logs. 
A small percentage of respondents used external tools for instrumentation.

An elaborate analysis of the survey answers is presented in Appendix I and a description of the solutions in Appendix J.

\emph{\textbf{Associations between solutions.}}
Fourth, we analysed the associations between the adopted solutions.
For this analysis, we modeled the adoption of solutions as dichotomous variables and  analysed the Phi coefficient. 
To determine the statistical significance of the observed associations, we performed Chi-Square tests with a significance level of $0.05$. 
We found multiple significant medium to strong associations ($\phi>0.4$), of which we report an illustrative selection.
We also analysed the associations between solutions using the Jaccard similarity, which entails analogous results.

\begin{table}[t]
    \caption{Solutions mapping to the ISO/IEC 25010 model.}
    \label{tbl:tactics_maapping}
    \centering
    \begin{tabular}{p{6em}p{12.5em}p{8.5em}}
        \toprule
        
        \textbf{Characteristics} & \textbf{Sub-characteristics} &  \textbf{Solutions} \\ 
        \midrule
        
        \multirow{2}{*}{Func. Suitability} & Func. Completeness & 1, 2, 3  \\ 
                                           & Compliance & 3 \\
        \midrule
        Performance efficiency & Capacity & 2, 11  \\
        
        \midrule
         \multirow{2}{*}{Compatibility} & Co-existence &  7, 8 \\ 
                                        & Interoperability & 6, 12   \\

        \midrule
         Usability & User Error Protection & 19, 20 \\

        \midrule
        \multirow{2}{*}{Reliability} & Availability & 10, 16, 18  \\ 
                                     & Fault Tolerance & 9, 10 \\ 
        
        \midrule
        Security & Accountability & 19  \\ 
        
        \midrule
        \multirow{5}{*}{Maintainability} & Modularity &  4, 5, 7, 8, 12 \\ 
                                         & Reusability &  4, 5, 6, 8 \\ 
                                         & Analysability & 11, 17  \\ 
                                         & Modifiability & 4, 5 \\ 
                                         & Testability &  5, 7, 9, 10, 11, 14, 15  \\ 
        \midrule
        \multirow{5}{*}{Portability} & Adaptability (with Scalability) & 13, 16, 18  \\ 
                                     & Deployability (with Installability and Replaceability) & 12, 16, 17, 18  \\
    
        \bottomrule
    \end{tabular}
\end{table}

%

For example, designing separate modules/services for data quality assessment (5) is associated with the the design of independent modules/services in component coupling (7) ($\phi=0.43$).
Moreover, the use of one communication middleware to reduce coupling (7) is associated with standardisation and reuse of model interfaces between training and serving (6) ($\phi=0.51$).
Design of independent modules/services in component coupling (7) is also associated with CI/CD in maintenance of \ac{ML} component (16) ($\phi=0.54$).
These results indicate the solutions may be complementary and  suggest their joint adoption can be interdependent and incremental.

Similarly, model tests (15) are associated with data tests (15) ($\phi=0.63$) and  
integration tests (15) are associated with test automation (15) ($\phi=0.88$).
However, \ac{ML} tests (\eg~data tests) are not associated with traditional software tests (\eg~unit tests) or with test automation.
These results indicate a separation between \ac{ML} and \ac{SE} concerns exists.
Moreover, they indicate that mature teams jointly adopt advanced test practices, as also noticed in~\cite{breck2017ml}.

\emph{\textbf{Architectural styles for \ac{ML}.}}
Next, respondents were asked to select the architectural styles employed in their projects.
The results are illustrated in Fig.~\ref{fig:survey_arch_tyles} and  indicate that the majority of respondents used the event-driven style.
Nonetheless, the difference between event-driven, lambda and  micro-service/SOA architecture styles is not large.
Although we did not find significant associations between the architectural styles, the lambda architecture can be used concomitantly with other architectural styles.
We also searched in literature for evidence to strengthen these findings, but could not find any study on this topic.
 
\emph{\textbf{Quality attributes.}} Last, respondents were asked to link the solutions to software quality attributes (characteristics) from ISO/IEC 25010~\cite{iso25010}, which enabled to restate them as architecture tactics~\cite{bass2003software}.
Tactics are architectural building blocks from which design patterns can be created and  represent architectural decisions that improve individual quality attributes~\cite{bass2003software,harrison2010architecture}. 
Therefore, the results of this analysis provide direct guidance for practitioners who aim to improve specific quality attributes of systems with \ac{ML} components.

Since the solutions do not presume a ranked order, we considered all solutions equally important.
The final results are presented in Table~\ref{tbl:tactics_maapping}.
We note that ``Scalability" and ``Interoperability"~--~considered important decision drivers (Fig.~\ref{fig:survey_decision_drivers})~--~are addressed by multiple solutions.
Similarly, ``Maintainability", considered to have the biggest impact on \ac{SA} by interview participants (Fig.~\ref{fig:impact_interviews}), is addressed by the largest number of solutions.
We also observe that some quality attributes from the standard (\eg~Operability or Maturity) are not addressed by any solution and  note this result does not imply that missing quality attributes are not challenging.
Instead, some quality attributes may not be applicable or require adaptation to accommodate \ac{ML} components, as previously suggested by~Kuwajima et al.~\cite{kuwajima2020engineering}.

We also mention that the ``Compliance" sub-characteristics are not present in the quality standard because compliance is considered part of the overall system requirements.
Therefore, ``Compliance" spans all characteristics in Table~\ref{tbl:tactics_maapping}.
To avoid confusion, we represent ``Compliance" as a sub-characteristic of ``Functional Suitability".
\section{Discussion}
\label{sec:discussion}

We comment on several aspects of the study.
First, regarding the challenges discovered from interviews, we analysed the percentage of survey respondents who did not have a strategy to tackle them.
We found out that \numprint{23}\% of respondents had no strategy for challenge (19) and  \numprint{22}\% of respondents had no strategy for challenge (20). 
These results show that more than 75\% of respondents tackled these challenges and  bring evidence that both challenges are relevant, in spite of the fact that they were mentioned in one interview each.
Moreover, the answers for challenge (20) have similar distributions for teams consisting of 6-9 and 10-15 members, suggesting the challenge is not motivated by team size.

Second, both in the interviews and in the survey, the architectural decision drivers for trustworthy \ac{ML}~\cite{EUGuidelines}~--~\ie~``Robustness", ``Security", ``Privacy"~--~were not considered critically important by respondents.
McGraw et al.~\cite{mcgraw2020architectural} argue that, from a security engineering perspective, the \ac{SA} of systems with \ac{ML} components is an important first step.
However, the results of our study show that practitioners focus on lower level concerns, such as performance or scalability.
Similar data was also observed previously~\cite{serban2021practices}.


Third, we noticed that some solutions have low adoption.
For example, self-adaptation for managing inherent uncertainty (9) was used by less than 5\% of respondents.
While this result may seem surprising, according to
Mahdavi-Hezavehi et al.~\cite{mahdavi2017classification} the number of self-adaptation techniques for ``automated learning'', \ie~\ac{ML} components, is small.
We conjecture that a small number of solutions are not applied because they are not fit, or still prototypes in academia.

Last, when mapping the solution to software quality attributes, we used the mature and authoritative ISO/IEC 25010 standard~\cite{iso25010}.
At the moment, no similar model exists for \ac{ML} components, although policy makers indicate such models are under development~\cite{jobin2019global}.
The results from our study indicate that, with the exception of ``Interpretability",  \ac{ML} specific quality attributes are not yet critical \ac{SA} decision drivers.
However, we expect this to change once mature quality models for \ac{ML} are available.

\emph{\textbf{Threats to validity.}}
We identified three potential threats to validity, corresponding to the three stages of the study.
First, the \ac{SLR} can be affected by missing or exclusion of relevant papers.
To mitigate this threat, we used multiple digital libraries for information retrieval.
Additionally, we complemented the results with grey literature (manual search) and  through snowballing.
The researchers' bias in the data extraction was prevented using a data extraction form (Appendix C), which allowed consistency in data analysis and  through discussions between authors with different backgrounds.

Second, the data from interviews may be subject to bias.
To limit this bias, we analysed the participants' profiles and ensured they have relevant experience for the study.
We recruited participants with diverse backgrounds and experience, working for organisations with distinct sizes and experience in \ac{ML}.
We also used two strategies to alleviate memory bias, \ie~we shared a short version of the interviews before the meeting and  asked participants to share their experience from a recent project in the preliminaries.
Moreover, we assured participants of data confidentiality and anonymisation, in order to limit participants from answering the questions in a manner that would better position them. 

Third, to limit the survey bias we included additional fields besides the answers from \ac{SLR} or interviews (\eg~Other fields for all challenges).
We also advertised the survey to diverse groups, in order to limit selection bias. 
Nonetheless, as shown in Section~\ref{sec:survey}, some groups of respondents are under-represented and  may introduce selection bias.
This bias can be removed by gathering more data, as we plan to do in the future.
Last, to avoid researchers' bias, we used data triangulation from multiple sources.

\section{Conclusions and future research}
\label{sec:conclusions}

We studied how systems can be (re-) architected to enable robust adoption of \ac{ML} components. 
We ran a mixed-methods empirical study consisting of: (i) a \ac{SLR} which revealed \intRelevantArticles~relevant articles, from which we complied \intInitialChallenges~\ac{SA} challenges (and solutions) for \ac{ML}, (ii) \intParticipants~semi-structured interviews which revealed \intInterviewChallenges~new challenges and \intInterviews~new solutions and  (iii) a survey with \intSurveyParticipants~architects.

We reported on the impact of each challenge on \ac{SA} and  the main \ac{SA} decision drivers for \ac{ML}.
Moreover, we established a link between solutions and quality attributes from the ISO/IEC 25010 standard, which allowed us to provide practitioners with twenty architectural tactics for systems with \ac{ML} components. 

For future research we plan to increase the number of respondents to the survey and enable more robust analyses.
Moreover, we plan to add depth to the interpretation of the results through validation interviews and expand the quality attributes from ISO/IEC 25010 with \ac{ML} specific quality attributes. 

\bibliographystyle{ieeetr}
\bibliography{bibliography}



\end{document}